\documentstyle[12pt]{article}

\newfont{\msbm}{msbm10}

\def\g{\gamma}

\def\CC{{\cal C}}

\def\N{\hbox{\msbm N}}

\def\Z{\hbox{\msbm Z}}

\def\R{\hbox{\msbm R}}

\def\H{{\cal H}}

\def\m{\mu}

\def\l{\lambda}

\newtheorem{prop}{Proposition}
\newtheorem{cor}{Corolary}
\newtheorem{theo}{Theorem}
\newtheorem{deff}{Definition}

\def\bc{\begin{cor}}
\def\ec{\end{cor}}

\def\bt{\begin{theo}}
\def\et{\end{theo}}

\def\bd{\begin{deff}}
\def\ed{\end{deff}}

\def\bp{\begin{prop}}
\def\ep{\end{prop}}

\def\ba{\begin{eqnarray}}
\def\ea{\end{eqnarray}}

\def\be{\begin{equation}}
\def\ee{\end{equation}}

\newfont{\msbms}{msbm6}  

\def\Zi{\hbox{\msbms Z}}

\def\mo{\mbox{$\mu_0$}}

\def\rep{representation}
\def\reps{representations}

\begin{document}


\title{
Comments on the kinematical structure of\\ loop quantum cosmology}

\author{J.\ M.\ Velhinho}


\date{{\it Departamento de F\'\i sica, Universidade da Beira 
Interior\\R. Marqu\^es D'\'Avila e Bolama,
6201-001 Covilh\~a, Portugal}\\{jvelhi@dfisica.ubi.pt}}

\maketitle

\begin{abstract}

\noindent  
We comment on the presence of spurious observables and on a subtle
violation of irreducibility in loop quantum cosmology.
\end{abstract}


\pagestyle{myheadings}


\noindent
\subsubsection*{Introduction}
Loop quantum cosmology (LQC) originated from the works of M.~Bojowald (see e.g.~\cite{BMT} and references therein)
and was later on reformulated in~\cite{ABL}. It is an application
to quantum cosmology of methods and ideas  adapted from those used in loop quantum 
gravity\footnote{For reviews of loop quantum gravity see~\cite{T1,R2,AL}.} (LQG).
This adaptation of LQG methods to the quantization of classical cosmological models produced 
spectacular results (see e.g.~\cite{B1,BMT}). However,
foundational and conceptual issues are still open.
 
LQC is in several aspects quite different from LQG, and presents difficulties
that are absent in LQG. These are mainly due to the lack of diffeomorphism invariance, 
which plays a crucial role in the very construction and methods of LQG.

An important open question is the appearance of a large surplus of solutions of the LQC Hamiltonian 
constraint~\cite{ABL,B1}. Partly related to this is the nonseparability of the kinematical
Hilbert space, i.e.~it does not possess a countable basis. 
(Here kinematical means that the Hamiltonian constraint is still
to be imposed.) Nonseparability, eventhough at the kinematical level, is an odd feature in quantum mechanics.
Moreover, there are increasing indications that even in full LQG a separable
Hilbert space is achieved prior to imposing the Hamiltonian constraint~\cite{Z,R2,T3,V,FR}.

We will try to show in the  present letter that nonseparability is not mandatory in LQC.
Moreover, it is our understanding that this question is surpassed in importance by two intrinsic aspects
of the current LQC formulation.
The first question concerns a 
subtle violation of irreducibility. The second is an unexpected commutation involving the 
quantum Hamiltonian, 
which is responsible for the appearance of spurious observables, and therefore for the excess of solutions
of the Hamiltonian constraint. 
Both questions are  related to  the particular form of the
LQC Hamiltonian constraint operator.

We  assume in what follows that the reader is familiar with the current formulation 
of LQC as given in~\cite{ABL} (see also~\cite{AL}), 
which we take as our starting point. Only a minimal summary of the necessary kinematical setup 
is included here.
As in~\cite{ABL}, we  restrict ourselves to the simplest situation of spatially flat, homogeneous
and isotropic cosmological models.

\subsubsection*{Kinematical setup}
The classical gravitational phase space 
that results from the connection variables formalism is 
coordinatized by real variables $c$ and $p$, with canonical Poisson bracket
\be
\label{ccr}
\{c,p\}=8\pi\g G/3,
\ee
where $\g$ is the Barbero--Immirzi parameter. The variable $p$ has dimensions of (lenght)$^2$
and is directly related to the scale factor $a$, $|p|=a^2$. The connection related variable
$c$ is dimensionless. There remains a single constraint, namely the Hamiltonian constraint:
\be
\label{h}
-{6\over \g^2}\, c^2\, {\rm sgn}(p)\, |p|^{1/2}+8\pi G\, C_{\rm matter}=0,
\ee
where sgn is the sign function and $C_{\rm matter}$ is the matter Hamiltonian.

The kinematical variables chosen in~\cite{ABL} to quantize (the gravitational part of) this system 
consist of $p$ and 
the commutative  algebra
$\CC$ of almost periodic functions of $c$, whose elements are finite linear combinations
of exponentials $e^{i\m c/2}$, where $\m$ can take any real value. The algebra $\CC$
clearly separates points on $\R$. Together with $p$, we have a Poisson algebra, with
\be
\label{eccr}
\{e^{i\m c/2},p\}={8i\pi\g G\m\over 6} e^{i\m c/2},
\ee
which is complete, i.e.~it separates points in phase space.

The LQC quantization  proceeds as follows. Let $\H$ be the
Hilbert space spanned by mutually orthogonal vectors $|\m\rangle$, $\m\in\R$,
$\langle\m'|\m\rangle=\delta_{\m'\m}$, where $\delta_{\m'\m}$ is the Kronecker delta
(not the Dirac distribution). By construction, $\H$ is nonseparable.
The quantum operators $\hat p$ and $\widehat {e^{i\m c/2}}$ are defined by
\be
\label{phat}
\hat p|\m\rangle={8\pi\g G\hbar\over 6}\m|\m\rangle={\g \ell_{P}^2\over 6}\m|\m\rangle,
\ee
where $\ell_{P}=(8\pi G\hbar)^{1/2}$ is the Planck lenght, and 
\be
\label{ehat}
\widehat {e^{i\m' c/2}}|\m\rangle=|\m+\m'\rangle.
\ee
The commutation relations
\be
\label{qeccr}
[\widehat {e^{i\m c/2}},{\hat p}]=i\hbar {8i\pi\g G\m\over 6} \widehat{e^{i\m c/2}}
\ee
are satisfied and the \rep\ is irreducible.

Let us consider the gravitational part of the Hamiltonian:
\be
\label{hg}
C_{\rm grav}=-{6\over \g^2}\, c^2\, {\rm sgn}(p)\, |p|^{1/2}.
\ee
The LQC quantum Hamiltonian constraint operator 
$\hat C_{\rm grav}$ is given by:
\be
\label{qh}
\hat C_{\rm grav}|\m\rangle=3(\g^3\mu_0^3\ell_{P}^2)^{-1}(V_{\m+\mu_0}-V_{\m-\mu_0})
(|\m+4\mo\rangle+|\m-4\mo\rangle-2|\m\rangle),
\ee
where $V_{\m}:=(\g|\m|/6)^{3/2}\ell_p^3$ and \mo\ is a  regularization parameter.
Effectively, $\m_0$ parametrizes a quantization ambiguity, as the regulator cannot be removed.
The currently adopted viewpoint~\cite{ABL,AL} is that this ambiguity can be naturally removed
by invoking results from LQG and that (\ref{qh}), with the parameter set to $\m_0=\sqrt 3/4$, 
is the correct Hamiltonian constraint operator. 

The kinematical setup is completed with the construction of the quantum inverse scale factor
(the quantum scale factor being already determined by $\hat p$).
The explicit form of this operator is not needed in what follows; it is sufficient to say
that, like the quantum scale factor, the quantum inverse scale factor commutes with $\hat p$,
and therefore its eigenvectors are the basis vectors $|\mu\rangle$.
\subsubsection*{Reducibility}
Here, and throughout this work, we consider only \reps\ such that zero belongs to the 
spectrum of $\hat p$.

Let us define $\l_k:={{4\m_0}\over {k+1}}$, $k=0,1,2,\ldots$, and let $\xi$ be a real number such that $\mo/\xi$ is
irrational. Let $\CC(\lambda_k,\xi)$ denote the subalgebra of $\CC$ 
generated by
the two exponentials $e^{i\lambda_k c/2}$ and $e^{i\xi c/2}$, 
\be
\label{cle}
\CC(\lambda_k,\xi):=\{\sum_{n,m\in\Zi} a_{nm}
e^{{i\over 2}(n\lambda_k+m\xi)c}\}.
\ee
Since $\l_k/\xi$ is irrational, it is clear that $\CC(\lambda_k,\xi)$ separates points on $\R$.
Thus, the Poisson subalgebras defined by $p$ and any given algebra $\CC(\lambda_k,\xi)$ are 
classically complete. Moreover, every algebra $\CC(\lambda_k,\xi)$ contains proper subalgebras of the same 
type, namely $\CC(\lambda_k,N\xi)$, where $N$ is a nonzero integer. The corresponding Poisson subalgebras
are therefore also complete.
All these Poisson algebras are, of course, unambiguously quantized in the Hilbert space $\H$.

Let $\H_k^{\xi}$ be the closed subspace of $\H$ spanned by vectors of the form
$|n\l_k+m\xi\rangle$, where $n,m\in\Z$.
Each space $\H_k^{\xi}$ is obviously a separable proper
subspace of $\H$. Although $\H_k^{\xi}$ does not carry a \rep\ of $\CC$,
it carries a (Dirac) \rep\ of the kinematical algebra defined by $\CC(\lambda_k,\xi)$
and $p$. We have thus found a  complete Poisson
algebra, in fact a large family of them, that is quantized but whose action on $\H$ is not 
irreducible.

In the usual quantization of unconstrained systems the latter conclusion would be, we believe,
a clear indication that the \rep\ space $\H$ is too big
(see a discussion in this respect in~\cite{G}). In the presence of constraints the question becomes
more subtle, as one has to make sure that the quantum constraint operators are not lost when
trying to reduce the \rep.

In the present case, it is obvious that the only constraint operator, namely $\hat C_{\rm grav}$ (\ref{qh}),
has an action on every space  $\H_k^{\xi}$. The quantum
scale factor and the quantum inverse scale factor operators act within $\H_k^{\xi}$ as well.
Thus, a complete kinematical setup descends from $\H$ to each of the separable spaces $\H_k^{\xi}$. 
We see no reason  to work with the full algebra $\CC$, and are therefore led to conclude
that the space $\H$ is unnecessarily big. Moreover, for $\lambda_k$ of the form $\m_0/n$, 
i.e.~for $k+1=4n$, $n\in\N$, one can forget $\H$ altogether, start from $\H_k^{\xi}$,
and reproduce step by step the whole quantization process described in~\cite{ABL}. Special 
values of $k$ are required for this purpose since, for instance, the construction of the quantum Hamiltonian
uses the operator $\widehat {e^{i\mu_0 c/2}}$, which is not defined in $\H_k^{\xi}$
for arbitrary $k$.

It turns out that the "obvious" solution to the above reducibility problem, and simultaneously to the
separability issue, namely to drop
$\H$ and continue in some space $\H_k^{\xi}$,
is not at all straightforward. Besides ambiguity in the choice of such a space, a major
concern is that every such \rep\ is again affected by reducibility. As mentioned
above, the Poisson algebra associated to $\CC(\lambda_k,N\xi)\subset\CC(\lambda_k,\xi)$ 
is still complete. Thus, one should consider the proper subspace 
$\H_k^{N\xi}\subset\H_k^{\xi}$
and reduce the representation again. It is clear that there is no end
to this process.

\subsubsection*{Spurious observables}
A known open question in the current LQC formulation is the appearance of an uncountable surplus of
solutions  of the quantum Hamiltonian constraint~\cite{ABL,B1}.
This is obviously reduced to an infinite countable set if one of the $\H_k^{\xi}$ \reps\ is used,
but the problem remains essentially the same.

The space of solutions is the (generalized) eigenspace of the Hamiltonian operator
corresponding to the (generalized) eigenvalue zero. The dimensionality of this space is dictated
by the set of operators commuting with the quantum Hamiltonian, i.e.~the set of observables.
The appearance of spurious solutions therefore reflects problems with the \rep\ of true observables,
or the presence of spurious observables, i.e.~operators that commute with the quantum Hamiltonian
but actually should not. We believe that spurious observables are  present in LQC.

In fact, it is  
trivial to check that $\hat C_{\rm grav}$ (\ref{qh}) commutes with the  operator
\be
\label{V}
V({4\m_0}):=\exp\left({6i\over {\g\ell_p^2}}{2\pi\over {4\m_0}}\hat p\right) 
\ee
(and with $(V(4\m_0))^N$, $\forall N\in\Z$). This result remains true after inclusion of 
the matter 
Hamiltonian, since, apart from matter degrees of freedom, it contains only the quantum scale
factor and the quantum inverse scale factor, and both these operators commute with $\hat p$.
The former commutation  apparently  has no classical correspondence.
Moreover, if one removes by hand the degeneracy in the solution space corresponding to the operator
$V(4\m_0)$, namely by selecting the subspace where
the (dual of the) operator acts trivially (i.e.~as the identity operator), one recovers exactly
the expected number of solutions in the gravitational sector, namely two (in the flat, homogeneous
and isotropic case under consideration).

\subsubsection*{Final remarks}
Both the reducibility issue and the appearance of spurious observables, and therefore the surplus of solutions
of the Hamiltonian constraint, are due to the presence of the regulator $\m_0$ in $\hat C_{\rm grav}$ (\ref{qh}).
Unless the ambiguity in the quantization of the Hamiltonian is handled in a totally different way, 
the appearance
of such a fixed parameter seems unavoidable. Moreover, in the currently accepted point of view, the parameter $\m_0$ 
is actually fundamental, i.e.~it is determined by the fundamental theory, namely LQG, that expectedly supports the
effective  LQC approach.

How can one then interpret the difficulties created by the special form of the LQC quantum Hamiltonian?
One possible  interpretation, obviously biased by Bojowald's original approach, is to accept that  the 
operator $V(4\m_0)$, or the operator 
$V(\m_0):=(V(4\m_0))^4$, 
has no meaning whatsoever, not even kinematically. Although radical, this hypothesis finds some support 
in the form of 
$\hat C_{\rm grav}$ itself. One such formal indication is that the spectrum of $\hat p$ is \rep-dependent,
the only invariant feature being precisely the  spectral points of the form
$4n\m_0\g\ell^2_P/6$, or $n\m_0\g\ell^2_P/6$, if one restricts attention to spaces 
$\H_k^{\xi}$ with $k+1=4m$.

The above attitude naturally leads to a strong condition on the quantum 
kinematical Hilbert space, namely that the action of
the operator $V(4\m_0)$, or $V(\m_0)$, should be trivial.
Trivialization of the operator $V(\m_0)$ leads to the kinematical Hilbert space used in~\cite{B2}.
Trivialization of $V(4\m_0)$ leads to a subspace of the later space, where the remaining
four-fold degeneracy is also removed. In this case, however, one would still have to rely on the
bigger space in order to construct the quantum Hamiltonian.


\subsubsection*{Acknowledgements}
\noindent It is a pleasure to thank Jos\'e Mour\~ao  and Thomas Thiemann.
This work was supported in part by 
POCTI/33943/MAT/2000, CERN/P/FIS/43171/2001 and POCTI/FNU/\-49529/2002.






\end{document}